\documentclass{PoS}

\title{Acceleration of Cosmic Ray Electrons at Weak Shocks in Galaxy Clusters}

\ShortTitle{Electron Acceleration at Cluster Shocks}

\author{\speaker{Hyesung Kang}\\
        Pusan National University\\
        E-mail: \email{hskang@pusan.ac.kr}}

\author{Dongsu Ryu\\
        UNIST\\
        E-mail: \email{ryu@sirius.unist.ac.kr}}

\author{T. W. Jones\\
        University of Minnesota\\
        E-mail: \email{twj@umn.ac.kr}}

\abstract{According to structure formation simulations, weak shocks with typical Mach number, $M_{\rm s}\lesssim 3$, are expected to form in merging galaxy clusters. The presence of such shocks has been indicated by X-ray and radio observations of many merging clusters. In particular, diffuse radio sources known as radio relics could be explained by synchrotron-emitting electrons accelerated via diffusive shock acceleration (Fermi I) at quasi-perpendicular shocks. Here we also consider possible roles of stochastic acceleration (Fermi II) by compressive MHD turbulence downstream of the shock. Then we explore a puzzling discrepancy that for some radio relics, the shock Mach number inferred from the radio spectral index is substantially larger than that estimated from X-ray observations. This problem could be understood, if shock surfaces associated with radio relics consist of multiple shocks with different strengths.
In that case, X-ray observations tend to pick up the part of shocks with lower Mach numbers and higher kinetic energy flux, while radio emissions come preferentially from the part of shocks with higher Mach numbers and higher cosmic ray (CR) production. We also show that the Fermi I reacceleration model with preexisting fossil electrons supplemented by Fermi II acceleration due to postshock turbulence could reproduce observed profiles of radio flux densities and integrated radio spectra of two giant radio relics. This study demonstrates the CR electrons can be accelerated at collisionless shocks in galaxy clusters just like supernova remnant shock in the interstellar medium and interplanetary shocks in the solar wind.}

\FullConference{35th International Cosmic Ray Conference – ICRC217-\\
		10--20 July, 2017\\
		Bexco, Busan, Korea}

\begin{document}

\section{Introduction}

Cosmological hydrodynamic simulations have shown that shock waves form due to supersonic flows in
the intracluster medium (ICM) during the structure formation in the Universe \cite{ryu03,vazza09}.
The evolution of such shocks indicates that shock surfaces behave like spherical bubbles blowing out from the cluster center
and could last for the dynamical time scale of clusters, $t_{\rm dyn}\sim 1$~Gyr.
This implies that shock bubbles are likely to expand out well beyond the virial radius of galaxy clusters.
Figure 1 illustrates that the supersonic flows penetrating deeply in the ICM along neighboring filaments could induce shocks near the virial radius, while accretion flows infalling onto the cluster induce strong accretion shocks several Mpc away from the cluster center \cite{hong14}.

It is well accepted that cosmic ray (CR) particles are produced via the diffusive shock acceleration (DSA) 
process through their interactions with MHD waves in the converging flows across astrophysical shocks \cite{bell78}.
In the {\it test particle limit} where the CR pressure is dynamically insignificant,
it predicts that the CR momentum spectrum at the shock
with the sonic Mach number $M_{\rm s}$ has a power-law form, $f(p) \propto p^{-q}$,
where $q=4M_{\rm s}^2/(M_{\rm s}^2-1)$ \cite{dru83}.
Hence, the DSA model can provide a natural explanation for the spatially resolved
synchrotron radiation spectrum at the shock location, $j_{\nu}(x_s)\propto \nu^{-\alpha_{\rm sh}}$,
where the radio spectral index is $\alpha_{\rm sh} = (q-3)/2$.

\begin{figure*}[t]
\centering
\includegraphics[width=130mm]{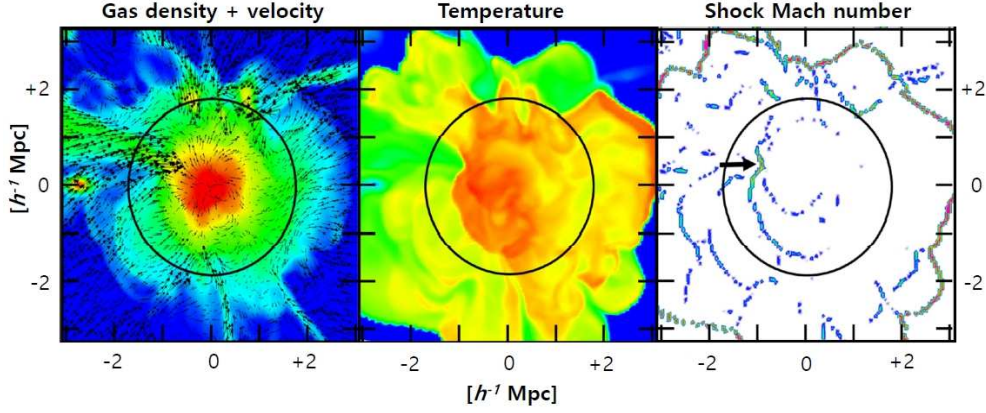}
\caption{Distributions of
the gas density and velocity vector (left), temperature (middle), and shock Mach number (right)
around a mock galaxy cluster in a structure formation simulation \cite{hong14}. The circle shows the virial radius, $r_{200}$, while the arrow points out the shock with the largest ram pressure, $\rho_1 u_s^2$. 
}
\label{fig1}
\end{figure*}

Diffuse radio structures known as the `radio relics' have been detected in many merging clusters, and they are interpreted as synchrotron emission from relativistic electrons accelerated at merger-driven shocks. Two famous giant radio relics are nicknamed as the `Sausage relic' \cite{vanweeren10} and the `Toothbrush relic' \cite{vanweeren12} because of their arc-like or linear shapes. Such elongated morphology is consistent with the spatial distribution of shocks near $r_{200}$ in Figure 1. Many of the observed features of these radio relics can be explained by DSA model:
(1) thin elongated shapes, 
(2) gradual spectral steepening from the relic edge toward the cluster center,
(2) power-law-like integrated spectrum with the spectral index, $\alpha_{\rm int}\approx \alpha_{\rm sh} + 0.5$,
and (3) high polarization levels up to 50\% that is expected from the shock compression of turbulent magnetic fields
\cite{vanweeren10,brunetti2014}. In order to study the origin of radio relics, we perform DSA simulations for CRe at structure formation shocks and compare the model predictions with the observational data of the aforementioned relics.

\section{Numerical Calculation}

\subsection{Numerical Modeling for DSA with Postshock Turbulent Acceleration}

We follow the electron acceleration via DSA (Fermi I) at 1D spherical shocks, and radiative cooling and turbulent (Fermi II) acceleration 
in the postshock region by solving the following diffusion-convection equation:
\begin{eqnarray}
{\partial g_{\rm e}\over \partial t} + u {\partial g_{\rm e} \over \partial r}
= {1\over{3r^2}} {{\partial (r^2 u) }\over \partial r} \left( {\partial g_{\rm e}\over
\partial y} -4g_{\rm e} \right) 
+ {1 \over r^2}{\partial \over \partial r} \left[r^2 \kappa(r,p){\partial g_{\rm e} \over \partial r} \right] \nonumber\\
+ p {\partial \over \partial y}\left[ {D_{pp} \over p^3} \left( {\partial g_{\rm e}\over \partial y} -4g_{\rm e} \right) \right] 
+ p {\partial \over {\partial y}} \left( {b\over p^2} g_{\rm e} \right),
\label{diffcon}
\end{eqnarray}
where $f_e(r,p,t)=g_e(r,p,t)p^{-4}$ is the pitch-angle-averaged phase space distribution function
for CRe, $u(r,t)$ is the flow velocity, $y=\ln(p/m_e c)$, $m_e$ is the electron mass, and $c$ is
the speed of light \cite{skill75}.
Here $r$ is the radial distance from the cluster center.

The electron energy loss term, $b(p)= \dot p_{\rm Coul} + \dot p_{\rm sync+iC} $, accounts for Coulomb scattering, 
synchrotron emission, and iC scattering off the cosmic background radiation.
We adopt a Bohm-like spatial diffusion coefficient, $\kappa(p)= \kappa_{\rm B} \cdot (p/m_ec)$ for relativistic electrons, 
where the normalization factor, $\kappa_{\rm B}= m_ec^3/(3eB)= 1.7\times 10^{19} {\rm cm^2s^{-1}}/B_{\mu G} $, 
with $B_{\mu G}$ expressed in units of $\mu G$.
Moreover, we explore a scenario in which the postshock electrons gain energy by Fermi II acceleration
via transit time damping (TTD) resonance with compressive fast-mode MHD turbulence \cite{brunetti2007}.
The momentum diffusion coefficient for TTD resonance can be modeled as
\begin{equation}
D_{pp} = { p^2 \over {4\ \tau_{\rm acc}}}, 
\label{Dpp}
\end{equation}
where $\tau_{\rm acc}$ is an effective acceleration time scale for turbulent acceleration \cite{kang17}. 
In order to model the decay of turbulence behind the shock, 
we assume the turbulent acceleration time increases behind the shock on the scale of $r_{\rm dec}$ as 
$\tau_{\rm acc} =\tau_{\rm acc,0} \exp\left[{(r_s-r)} /r_{\rm dec}\right]$, where 
$\tau_{\rm acc,0}\approx 10^8~yr$ and $r_{\rm dec}\approx 100$ kpc are adopted in the simulations considered here.

\subsection{Reacceleration Model}

Here we consider the reacceleration model, in which a shock propagates into the preshock region with preexisting low-energy CRe specified by the following power law spectrum with exponential cutoff :
\begin{equation}
f_{\rm pre}(p) = f_o \cdot p^{-s} \exp \left[ - \left({p \over p_{e,c}} \right)^2 \right],
\label{fpre}
\end{equation}
where the slope $s=4.6$ and the cutoff $\gamma_{e,c}=p_{e,c}/m_ec=300$ are adopted. The normalization factor, $f_o$, is adjusted by the ratio of the preexisting CRe pressure 
to the gas pressure in the preshock region, 
$N \equiv P_{\rm CRe,1}/P_{\rm gas,1} \propto f_o$.
Typically, the models with $N \sim 10^{-4}$ produce the radio flux levels that
can match the amplitude of observed flux densities for the two relic (see Table~1).

In the test particle regime for weak shocks, the reaccelerated spectrum of $f_{\rm pre}(p)$ at the shock 
can be calculated analytically by
\begin{equation}
f_{\rm reacc}(r_s, p)= q \cdot p^{-q} \int_{p_{\rm inj}}^p p^{\prime q-1} f_{\rm pre} (p^\prime) dp^\prime,
\label{freacc}
\end{equation}
where $p_{\rm inj}$ is the injection momentum above which CR particles participate in the Fermi I acceleration \cite{dru83}.
If the DSA slope, $q$, is smaller than the slope, $s$, of the preexisting population,
$f_{\rm reacc}(r_s, p)\propto p^{-q}$.
In that case, the only significant role of preexisting  
low-energy CRe is to provide seed particles to be injected to the DSA process.

\subsection{Shock Model Parameters}

\begin{table*}
\begin{center}
{\bf Table 1.}~~Parameters for Model Spherical Shocks\\
\vskip 0.3cm
\begin{tabular}{ lrrrrrrrrrrrr }
\hline\hline

Model & $M_{\rm X}$ &$M_{\rm radio}$ & $M_{\rm s,i}$ & $kT_1$ & $B_1$ & $t_{\rm obs}$& $M_{\rm s,obs}$ & $kT_{\rm 2,obs}$ & $u_{\rm s,obs}$& $N$\\
 {} & & & &{(keV)} &($\mu G$) & (Myr)& &{(keV)} & {(${\rm km~s^{-1}}$)}&($10^{-4}$)  & \\

\hline
Sausage &2.7 &4.6 & 4.0 &2.1& 1& 211 & 3.21  & 8.6 & $2.4\times10^3$ & $1.2$\\
\hline
Toothbrush  & 1.5 &2.8& 3.6 &3.0& 1& 144 & 3.03  & 11.2 & $2.7\times10^3$ & $5.0$\\
\hline
\end{tabular}
\end{center}
{$M_{\rm X}$: Mach number inferred from X-ray observations}\\
{$M_{\rm radio}$: Mach number estimated from observed radio spectral index at the relic edge}\\
{$M_{\rm s,i}$: initial shock Mach number at the onset of the simulations ($t_{\rm age}=0$)}\\
{$kT_1$: gas temperature in the preshock ICM}\\
{$B_1$: magnetic field strength in the preshock ICM}\\
{$t_{\rm obs}$: shock age when the simulated results match the observations}\\
{$M_{\rm s,obs}$: shock Mach number at $t_{\rm obs}$}\\
{$kT_{\rm 2,obs}$: postshock temperature at $t_{\rm obs}$}\\
{$u_{\rm s,obs}$: shock speed at $t_{\rm obs}$}\\
{$N=P_{\rm CRe}/P_{\rm g}$: the ratio of seed CR electron pressure to gas pressure in the preshock region}\\
{The subscripts 1 and 2 denote the states upstream and downstream of shock, respectively.}
\label{tab1}
\end{table*}

Here we assume that the shock dynamics can be approximated by a self-similar blast wave
that propagates through the isothermal ICM with the hydrogen number density profile of $n_{\rm H}\propto r^{-2}$.
So the shock radius and velocity evolves roughly as $r_s\propto t^{2/3}$ and $u_s\propto t^{-1/3}$,
respectively, where $t$ is the time since the point explosion for the spherical blast wave \cite{ryu91}.
Table~1 summarizes two sets of the model parameters that can represent the Sausage and the Toothbrush relic.
The second and third columns show the shock Mach number inferred from X-ray and radio observations, respectively \cite{stroe16,vanweeren16}.
The ICM temperature, $kT_1$, specifies the sound speed of the preshock gas, while the shock Mach number at the onset of the simulation, $M_{\rm s,i}$, determines the initial shock speed. 
In our simulations, the spherical shock slows down in time, while the postshock electrons advect downstream and cool radiatively behind the shock.
So we selected the best time of `observation', $t_{\rm obs}$, by comparing the predicted profiles of radio flux densities $S_{\nu}(R)$ and the integrated radio spectrum $J_{\nu}$ with the observed data of the two relics.

The shock properties at $t_{\rm obs}$ given in Table~1 are in reasonable agreement with the observations of the two relics. 
In the case of the Sausage relic the discrepancy between $M_{\rm X}\approx 2.7 $ and $M_{\rm radio}\approx 4.6$ is alleviated greatly by $M_{\rm s,o}\approx 3.2$.
But for the Toothbrush relic, $M_{\rm s,o}\approx 3.0$ is still higher than $M_{\rm X}\approx 1.5$ .
We note that a radio relic may be associated with multiple shocks, and X-ray observations tend to pick up shocks with lower $M_{\rm s}$ along a given line-of-sight \cite{hong15}.
So $M_{\rm X}$ inferred from X-ray observations could be lower than $M_{\rm radio}$ estimated
from radio spectral index.

\begin{figure*}[t]
\centering
\includegraphics[trim=2mm 3mm 0mm 2mm, clip, width=120mm]{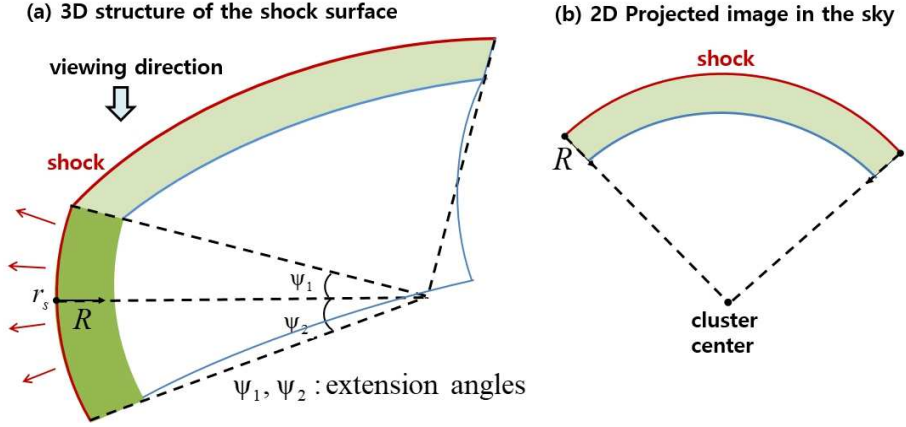}
\caption{
(a) Schematic diagram showing a spherical shock surface (red) and the postshock electron distribution (green) behind the shock. Viewing depth is defined by the extension angles $\psi_1$ and $\psi_2$.
(b) 2D image of a radio relic projected in the sky plane, if viewed from the top as indicated in (a).
}
\label{fig2}
\end{figure*}

We assume that the Sausage and the Toothbrush relic can be represented by a wedge-like patch of a spherical shell
with the radius $r_s$
shown in Figure 2, whose depth along the line-of-sight is specified by the extension angles $\psi_1$ and $\psi_2$.
Here $R$ is the downstream distance from the shock front.
The extension angles should be small, typically $\psi\sim 10^{\circ}$, in order to reproduce thin elongated images.
The inner radius of the spherical shell containing CRe is roughly defined by the electron cooling length, $u_s \cdot t_{\rm rad}(\gamma_e)$, where $t_{\rm rad}$ is the cooling time of electrons with the Lorentz factor $\gamma_e$.

In the case of the Sausage relic, the observed $J_{\nu}$ steepens above 2~GHz, which  
is inconsistent with a single power-law energy spectrum of relativistic electrons
that are expected to be accelerated by a steady planar shock \cite{stroe16}.
Such a spectral curvature could be explained,
if the relic is generated by the shock that sweeps through and moves out of a finite-size cloud with preexisting CRe \cite{kangryu16}.
So in the case of the Sausage relic we assume that the shock encounters a cloud of 
size $L_{\rm cloud}=624$~kpc containing $f_{\rm pre}(p)$, and then exits out of it at $t_{\rm exit}=200$~Myr.
Since the injection of CRe stops from $t_{\rm exit}$ to $t_{\rm obs}=211$~Myr,
the volume-integrated electron energy spectrum steepens toward high energies.
So the elapsed period of $(t_{\rm obs}-t_{\rm exit})\approx 10$~Myr controls the degree of the spectral curvature of $J_{\nu}$ in the model for the Sausage relic.

\section{Results}


\begin{figure*}[t]
\centering
\vspace{-0.0cm}
\includegraphics[trim=3mm 3mm 5mm 5mm, clip, width=140mm]{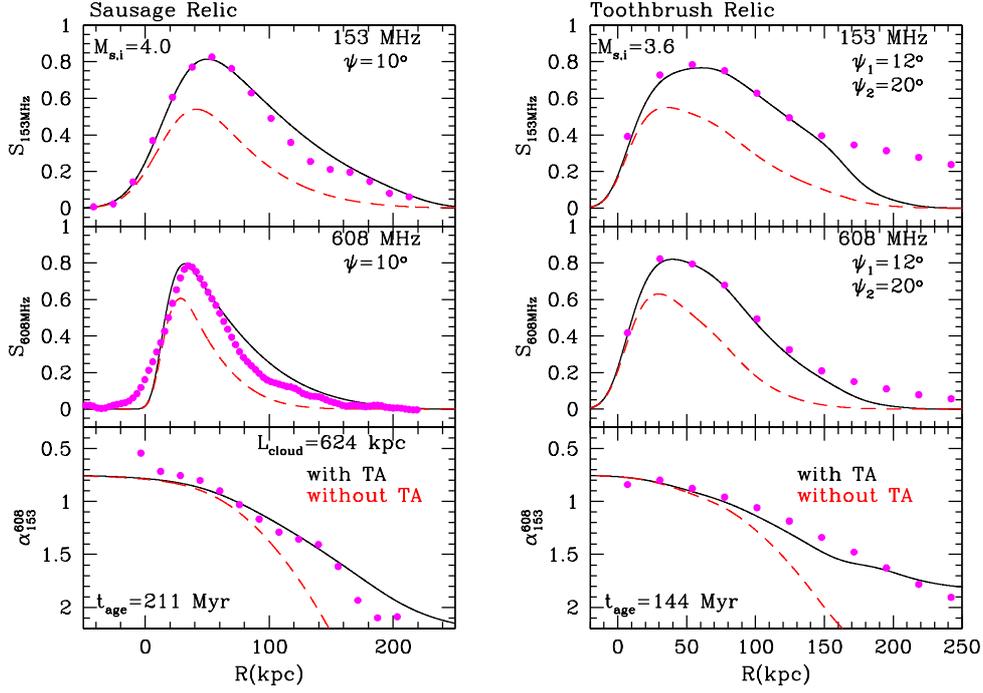}
\vspace{-4.5cm}
\caption{Radio flux density, $S_{\nu}$ at 150 MHz (top panels) and at 610 MHz (middle panels)
in arbitrary units, and the
spectral index, $\alpha_{150}^{610}$, between the two frequencies (bottom panels),
plotted as a function of the projected distance behind the shock, $R$ (kpc).
DSA models with (black solid lines) and without (red dashed lines) postshock turbulent acceleration are compared.
The magenta dots are the observational data from \cite{stroe16} for the Sausage Relic and from
\cite{vanweeren16} for the Toothbrush Relic.
}
\label{fig3}
\end{figure*}

Using the CRe energy spectrum and the magnetic field strength in the model DSA simulations,
we first calculate the synchrotron emissivity $j_{\nu}(r)$ of each spherical shell. 
The radio surface brightness, 
$I_{\nu}$, is calculated by adopting the wedge-like volume of radio-emitting electrons, specified with the
two extension angles relative to the sky plane, $\psi_1$ and $\psi_2$, as shown in Figure 2:
\begin{equation}
I_{\nu}(R)= \int_0^{h_{\rm 1, max}} j_{\nu}(r) d {\it h_1}+ \int_0^{h_{\rm 2, max}} j_{\nu}(r) d {\it h_2},
\label{SB}
\end{equation}
where $h_1 = r \sin{\psi_1}$ and $h_2 = r \sin{\psi_2}$ are the path lengths along line of sight beyond and in front of the sky plane, respectively. 

The predicted intensity, $I_{\nu}$, are convolved with relevant telescope beams in order to obtain mock radio flux density distribution, $S_{\nu}$.
Figure 3 demonstrates the mock flux profiles at $t_{\rm obs}$ are in good agreement with the observation data for the Sausage and the Toothbrush relic
and the Fermi II acceleration due to postshock turbulence plays a crucial role.


Figure 4 shows the time evolution of $\nu J_{\nu}$ for the two relics.
For the Sausage relic (left-hand panel), the red dashed line shows the spectrum at the first epoch just before the 
shock exits out of the cloud. 
Then the blue dot-dashed, black solid, magenta dot-long dashed, and green long dashed lines 
present the spectra with progressively steeper curvatures at four later epochs in chronological order.
In the case of the Toothbrush relic (right-hand) the integrated spectrum changes only slightly in time.
The black solid lines at $t_{\rm obs}$ are in reasonable agreement with the observed spectra in both relics.

\begin{figure*}[t]
\centering
\vspace{-0.0cm}
\includegraphics[trim=3mm 3mm 5mm 5mm, clip, width=140mm]{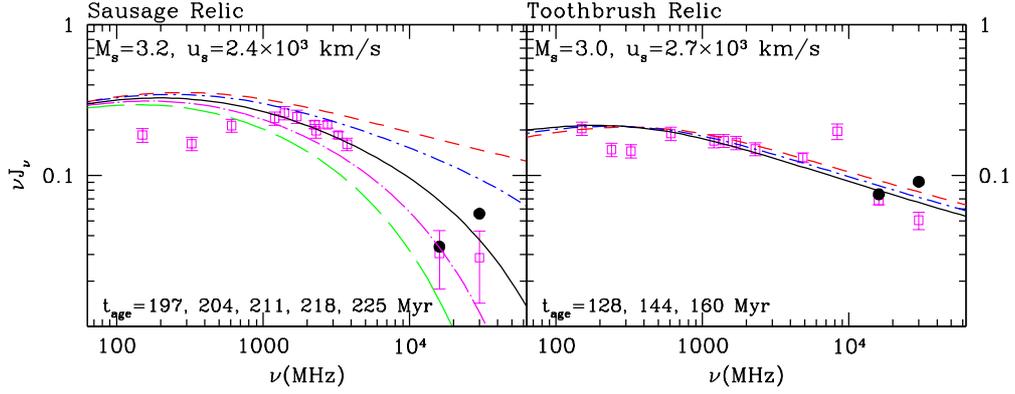}
\vspace{-8.8cm}
\caption{
Time evolution of volume-integrated radio spectrum at the shock ages
are shown in chronological order by the red dashed, green dot-dashed, black solid, magenta dot-long dashed, and blue long dashed lines.
The open magenta squares and the error bars are the observational data from \cite{stroe16}.
The solid black circles at 16 and 30 GHz are the data points, multiplied by factors of 1.11 and 1.96 for the Sausage relic and
by factors of 1.1 and 1.8 for the Toothbrush relic, respectively, which could represent the SZ corrected
fluxes \cite{basu16}.
}
\label{fig4}
\end{figure*}

\section{Conclusions}

Here we explore a reaccelration model in which a shock of $M_s\approx 3-4$ sweeps through a preshock 
cloud containing low-energy 
fossil electrons, and CR electrons are further accelerated by postshock turbulence via Fermi II process.
Transit time damping resonance off compressive MHD turbulence in the postshock flow is considered.
We find that turbulent acceleration with $\tau_{\rm acc}\approx 10^8$~yr is required
in order to match the observed broad profiles of the radio flux density, $S_{\nu}(R)$,
of the Sausage and the Toothbrush relic.

We demonstrate that our reacceleration model can reproduce the observed profiles of $S_{\rm 153MHz}$, $S_{\rm 608MHz}$,
and the spectral index $\alpha_{153}^{608}$ as well as the volume-integrated
spectrum $J_{\nu}$ of the Sausage radio relic \cite{stroe16} and the Toothbrush relic \cite{vanweeren16}.
The observed spectral curvature in $J_{\nu}$ of the Sausage relic can be explained by the finite size of the region containing fossil CRe.
The amount of fossil CRe required to explain the observed levels of $S_{\nu}$ is order of $10^{-4}$ of the gas pressure in the ICM.
The shock parameters for the best fitting models are given in Table~1.
This reacceleration scenario is consistent with the observational fact that
only a small fraction ($\sim 10\%$) of merging clusters host radio relics, since radio relics are generated only in the presence of fossil CRe.

This study confirms the scenario that CR electrons are accelerated at quasi-perpendicular ICM shocks in galaxy clusters, 
just like supernova remnant shocks in the interstellar medium.
Although CR protons are expected to be accelerated at quasi-parallel ICM shocks,
gamma-ray emission from galaxy clusters due to $\pi^0$ decay, which results from inelastic collisions between shock-accelerated CR protons and thermal protons, have not been detected so far \cite{ackermann14}.
This remains to be an important outstanding problem that needs to be investigated further through plasma simulations.

\section{Acknowledgments}
HK was supported by Basic Science Research Program through the National Research Foundation of Korea funded by the Ministry of Education (2014R1A1A2057940).
DR was supported by the National Research Foundation of Korea through grant 2016R1A5A1013277.
TWJ was supported by the US National Science Foundation through grant AST1211595.
The authors thank A. Stroe and R. J. van Weeren for providing the radio flux data for the Sausage relic in \cite{stroe16} and the Toothbrush relic in \cite{vanweeren16}, respectively.

\end{document}